# Creating RESTful APIs over SPARQL endpoints using RAMOSE

Marilena Daquino[a,b], Ivan Heibi[a,b], Silvio Peroni[a,b,*], David Shotton[a,c]

[a] *Research Centre for Open Scholarly Metadata, Department of Classical Philology and Italian Studies, University of Bologna, Bologna, Italy*

[b] *Digital Humanities Advanced Research Centre, Department of Classical Philology and Italian Studies, University of Bologna, Bologna, Italy*

[c] *Oxford e-Research Centre, University of Oxford, Oxford, United Kingdom*

**Abstract.** Semantic Web technologies are widely used for storing RDF data and making them available on the Web through SPARQL endpoints, queryable using the SPARQL query language. While the use of SPARQL endpoints is strongly supported by Semantic Web experts, it hinders broader use of RDF data by common Web users, engineers and developers unfamiliar with Semantic Web technologies, who normally rely on Web RESTful APIs for querying Web-available data and creating applications over them. To solve this problem, we have developed RAMOSE, a generic tool developed in Python to create REST APIs over SPARQL endpoints. Through the creation of source-specific textual configuration files, RAMOSE enables the querying of SPARQL endpoints via simple Web RESTful API calls that return either JSON or CSV-formatted data, thus hiding all the intrinsic complexities of SPARQL and RDF from common Web users. We provide evidence that the use of RAMOSE to provide REST API access to RDF data within OpenCitations triplestores is beneficial in terms of the number of queries made by external users to such RDF data using the RAMOSE API compared with the direct access via the SPARQL endpoint. Our findings show the importance for suppliers of RDF data of having an alternative API access service, which enables its use by those with no (or little) experience in Semantic Web technologies and the SPARQL query language. RAMOSE can be used both to query any SPARQL endpoint and to query any other Web API, and thus it represents an easy generic technical solution for service providers who wish to create an API service to access Linked Data stored as RDF in a conventional triplestore.

Keywords: REST API, OpenCitations, citation data, SPARQL endpoint, RDF, Linked Data, triplestore, RAMOSE, data access, query language

## 1. Introduction[1]

Application Programming Interfaces (APIs) are powerful means of automating communication between application programs and data services. The aim of an API is to expose service functions and data to facilitate the interaction with users or (particularly) machines. Representational State Transfer (REST)

APIs expose on the Web a set of stateless operations which enhance performance, reliability, and extensive reuse of the Web data resources [1].

Within the Semantic Web domain, the SPARQL 1.1 specifications include a Recommendation for "an application protocol for the distributed updating and fetching of RDF graph content in a Graph Store via the mechanics of the Hypertext Transfer Protocol (HTTP)"

---

[*]Corresponding author. E-mail: silvio.peroni@unibo.it.

[1] The first three authors contributed to the conceptualization of the work, the investigation, and the definition of the methodology. S. Peroni and M. Daquino developed the software described in Section 3 "RAMOSE: a technical introduction". M. Daquino is responsible for Section 4 "Use of REST APIs in Open Citations" (analysis and writing – original draft); I. Heibi is responsible for Section 5 "Related Works" (analysis and writing – original draft). All other sections were written and revised collaboratively by all the authors.

[2]. Such REST-based access to SPARQL endpoints has been a common ground used by several Semantic Web developers to query RDF data available on the Web [3]. Indeed, several institutions that have adopted Semantic Web technologies to manage their data – such as the British Library (http://bnb.data.bl.uk/), US government (https://www.data.gov/developers/semantic-web), and Wikidata (https://www.wikidata.org/) – usually employ such a REST-based approach to serve their RDF data to users (e.g. Web developers) and application programs via bespoke specialised Web interfaces that mediate the interaction with their SPARQL endpoints.

While SPARQL has widespread adoption among Semantic Web practitioners [4], it is not popular within the community of ordinary Web developers and scholars due to its complexity. The use of SPARQL is characterised by a very steep learning curve that prevents its widespread adoption in common Web projects, which usually leverage Web REST APIs to access and query data. Thus, the exclusive use of SPARQL endpoints to expose RDF data prevents easy access to such data by many stakeholders with legacy technologies. Indeed, several projects (including those of the institutions mentioned above) accompany their SPARQL endpoints with ad-hoc Web REST APIs. Such Web REST APIs are usually hardcoded and are difficult to maintain since they require expertise in both Web and Semantic Web technologies.

There is, thus, an increasing implicit demand for a generic mechanism that:

1. enables a broader Web audience (Web developers and scholars) to query RDF data available in triplestores behind SPARQL endpoint interfaces without having to use the SPARQL query language; and
2. allows Semantic Web developers to easily and quickly provide REST API access to their RDF data, a situation that we directly experienced in the context of OpenCitations [5] (https://opencitations.net/).

OpenCitations is an independent infrastructure organization for open scholarship dedicated to the publication of open bibliographic and citation data using Semantic Web technologies. Also, OpenCitations is engaged in advocacy for open citations – e.g. see the Initiative for Open Citations (I4OC, https://i4oc.org). Initially, the data within the OpenCitations Corpus [6] were queryable only by using our SPARQL endpoint. However, we received several suggestions from people working in different scholarly disciplines for a more holistic approach for data querying, to enable users with no skills in Semantic Web technologies to access these data and to reuse them for building Web applications. In addition to providing a standard Web REST API for access to our Corpus data, we also needed a method whereby we could quickly and easily create new Web REST APIs to extend such access to new RDF datasets that we ourselves might publish, while at the same time providing a generic tool for adoption by the Semantic Web community as a whole. Moreover, since our data sources interoperate with identifiers minted by external content providers (e.g. Crossref, ORCID, doi.org), we needed a flexible mechanism to serve RDF data integrated with information belonging to non-RDF data providers.

To address such needs, we developed RAMOSE, the RESTful API Manager Over SPARQL Endpoints (https://github.com/opencitations/ramose), which was explicitly created to foster reusability of RDF data across common Web applications. While developed to solve the specific problem of providing REST APIs for OpenCitations data, RAMOSE has been developed in a way which permits it to interact with *any* SPARQL endpoint, following the rationale we adopt for all our software development (available at https://github.com/opencitations), namely: while addressing the problem at hand, do this in a manner that provides a generic, open and broad tool which can be reused by others with similar requirements.

RAMOSE is an open-source Python software tool released under an ISC license. It allows one to create a Web REST API, with the related documentation, which acts as an interface to one or more SPARQL endpoints, regardless of the kinds of data hosted in the RDF triplestore. The creation of an API only requires the creation of a configuration file in a specific textual Markdown-like format (https://en.wikipedia.org/wiki/Markdown) which includes the SPARQL queries to be used by the API to retrieve RDF data. At OpenCitations, we now use RAMOSE to implement all the REST APIs introduced at http://opencitations.net/querying.

In this article, we provide a quick introduction to the context in which we have made this development (Section 2), followed in Section 3 by a description of RAMOSE, its architectural model, and how to configure and deploy it. In Section 4, we document how Web users engaged with OpenCitations data both before RAMOSE was developed and after we started using it to provide Web REST API access to OpenCitations datasets, to show the potential impact of RAMOSE on data reuse. After a discussion on other existing works concerning other software that addresses similar scenarios (Section 5), we conclude the article (Section 6) by sketching out some planned future developments.

## 2. Background: OpenCitations and its data

OpenCitations formally started in 2010 as a one-year project funded by JISC (with a subsequent extension). The project was global in scope and was designed to change the face of scientific publishing and scholarly communication, since it aimed to publish open bibliographic citation information in RDF [3] and to make citation links as easy to traverse as Web links. The main deliverable of the project, among several outcomes, was the release of an open repository of scholarly citation data described using the SPAR (Semantic Publishing and Referencing) Ontologies [7], and named the OpenCitations Corpus (OCC, http://opencitations.net/corpus), which was initially populated with the citations from journal articles within the Open Access Subset of PubMed Central (https://www.ncbi.nlm.nih.gov/pmc/tools/openftlist/).

At the end of 2015, we set up a new instantiation of the OpenCitations Corpus [6] based on a new metadata schema and employing several new technologies to automate the ingestion of fresh citation metadata from authoritative sources. From the beginning of July 2016, OCC started ingesting, processing and publishing reference lists of scholarly papers available in Europe PubMed Central. Additional metadata for these citations were obtained from Crossref (https://crossref.org) [8] and (for authors) from ORCID (https://orcid.org) [9]. Routine ingestion of new data into OCC from Europe PubMed Central was suspended in December 2017, when it contained 12,652,601 citation links. Since then, OCC has been used as a publication platform for citations derived from the ExCITE Project (http://excite.west.uni-ko-blenz.de/website/), the Venice Scholar Project (https://venicescholar.dhlab.epfl.ch/about) and other sources, and now contains 13,964,148 bibliographic citations to 7,565,367 cited publications.

Following our development in 2018 of Open Citation Identifiers (globally unique PIDs for citations treated as first-class data entities in their own right [10]), and using open references supplied by Crossref, we switched OpenCitations' bulk publication of citation links from OCC to COCI, the OpenCitations Index of Crossref open DOI-to-DOI citations [11], which was first released in July 2018 and currently contains 759,516,507 bibliographic citations between 60,778,357 DOI-identified publications. Also, in July 2018, in parallel with the development of COCI, we released the first version of RAMOSE and started to expose all OpenCitations data via Web REST APIs.

Recently, OpenCitations has been selected by SCOSS (https://scoss.org) among the Open Science infrastructures that deserve to receive funds from the community to foster their long-term sustainability.

## 3. RAMOSE: a technical introduction

RAMOSE is an open-source application written in Python which allows the agile development and publication of documented REST APIs for querying against any SPARQL endpoint. It is possible to customize RAMOSE to generate a Web REST API for the URL of a given SPARQL endpoint simply by creating an appropriate source-specific textual configuration file.

The modularity of RAMOSE allows a complete definition and customization of API operations and their input parameters. In addition, it enables one to apply pre-processing and post-processing steps by using external Python libraries that can be easily imported, and automatically generates HTML documentation of the API and a dashboard for the API monitoring.

RAMOSE has been designed to be consistent with the following principles:

1. It must work with any legacy RDF triple-store providing a public SPARQL endpoint.
2. A Semantic Web expert should only be required initially, to define the SPARQL queries hidden behind the API operations, while all the other aspects of the REST API configuration and use should not require Semantic Web skills.
3. API operations and their input parameters must be fully customizable according to the needs of the infrastructure exposing the data.
4. The configuration file of a RAMOSE-based API must be easy to write and must avoid technicalities as much as possible.
5. It must be possible to specify pre-processing and post-processing steps, developed as pure Python functions, in any operation, so as to better customise the interpretation of the input parameters and call outputs.
6. Basic built-in filters and refinement mechanisms must be provided by default.
7. It must be possible to use the REST API within another Python application, to run

it as a command line application, and to make it available as a proper service within a web server.

The source code of RAMOSE, its documentation, and examples of its use are all available on GitHub at https://github.com/opencitations/ramose. RAMOSE is licensed under the ISC License.

### 3.1. Architecture overview

RAMOSE is a middleware interface between the data consumer and one or more SPARQL endpoints. Figure 1 shows an overview of the application. It consists of the application file (i.e. the file `ramose.py`) and one or more configuration documents (one for each Web REST API service that is created by means of RAMOSE).

The RAMOSE application file handles the following aspects: service builder for running API operations, definition of built-in filters and refinement mechanisms, SPARQL query dispatcher, results format converter (either in CSV or JSON), generation of HTML documentation, and setting up of a web server for testing and monitoring purposes.

The application file accepts one or more configuration files in order to set up the services (one configuration file for each of the triplestores to be queried) and creates the APIs documentation and a web dashboard for testing and monitoring the API. Each RAMOSE configuration document contains metadata of one REST API service (name, contacts, license, description, etc.), the URL of the SPARQL endpoint to be queried, the optional specification of a Python file containing functions that can be used to pre-process the API call input, parameters and/or to post-process the result of the execution of the SPARQL query, and the definition of all the operations. Each operation must specify the SPARQL query to run against the SPARQL endpoint, the URL to call the operation, which includes also the name and shape of its input parameters, the HTTP method to use for the request, optional pre-processing/post-processing function names (defined in external libraries) to be executed before/after the execution of the SPARQL query, the types of the fields returned by the operation, an additional description of the operation, an example of use, and an exemplar output in JSON.

As shown in Figure 2, every time someone executes an operation, the URL of the call is parsed, and the values of the input parameters are retrieved according to the shape (i.e. data type and textual form) specified in the configuration file. The pre-processing functions are executed on the specified input parameters. These functions are used to transform the input parameters into a form appropriate for the SPARQL query to be executed.

Following this pre-processing, any input parameter included in the SPARQL query of the operation between `[[...]]` is replaced with its current value, and finally the SPARQL query is performed against the SPARQL endpoint according to the HTTP method specified. When the SPARQL endpoint returns a result, RAMOSE runs the post-processing functions on it. Post-processing allows one to perform simple tasks, such as data cleansing or normalisation, or more sophisticated processes, such as cross dataset queries for integrating results. After post-processing, RAMOSE applies filters and refinements (if specified in the call URL) and converts the results either into CSV or into JSON according to what has been specified in the request. Where JSON is chosen as the output format, it is possible to ask RAMOSE, via a particular refinement parameter included in the call URL, to transform the default JSON output into a more structured one. Refinement operations avoid operations generally performed via SPARQL (e.g. sorting results) that might affect the performance of the triplestore [12]. Moreover, these allow users to perform further operations on the results (since they cannot modify the underlying SPARQL queries). An example of the whole process is presented in Figure 2.

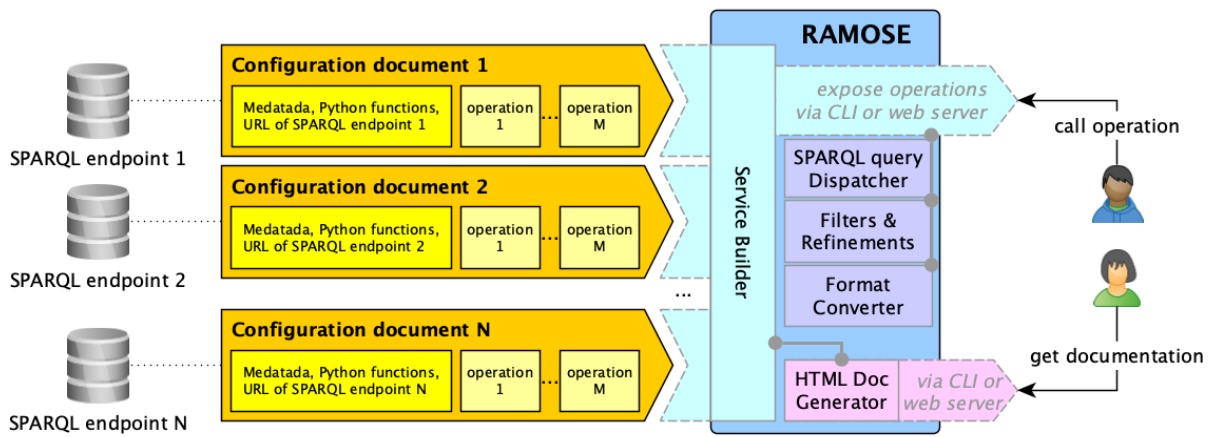

Figure 1. An overview of the main components of RAMOSE.

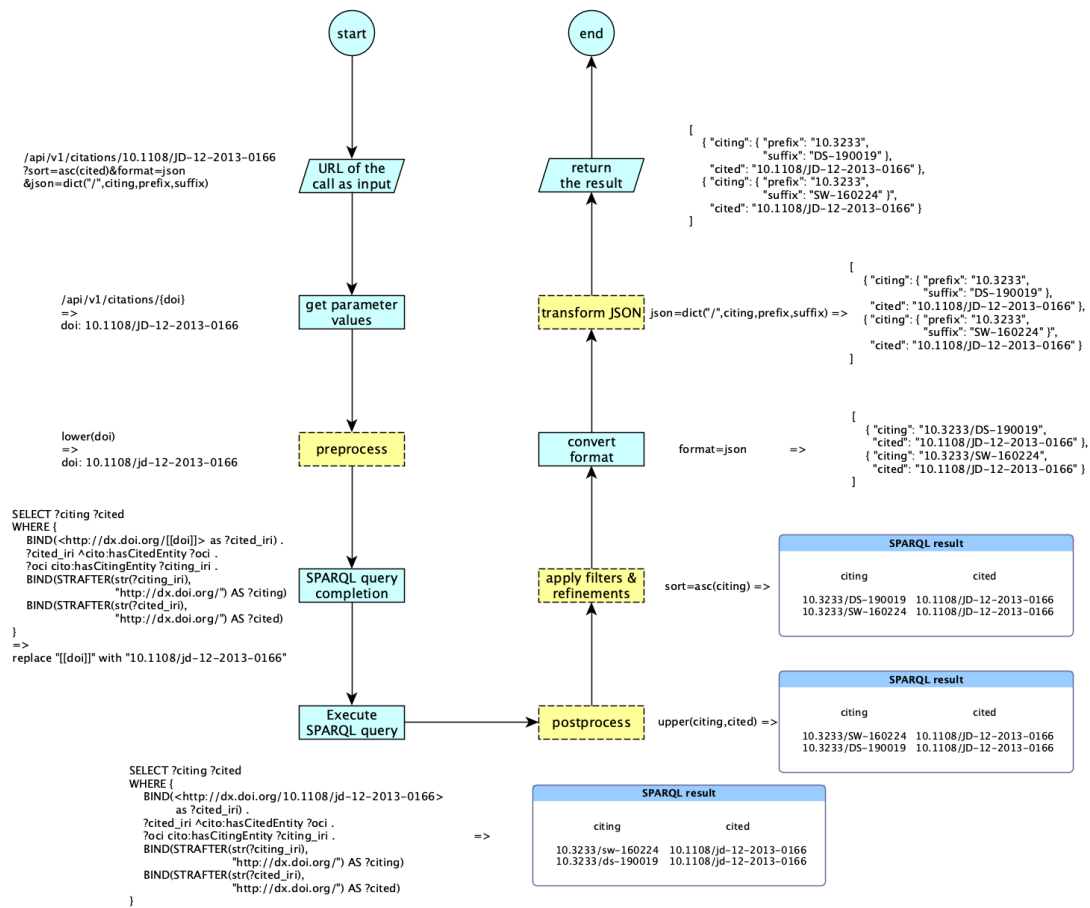

Figure 2. The workflow implemented by RAMOSE to handle an API call specified via a URL, accompanied by a running example. The yellow dotted rectangles are optionally executed since they depend on the call URL (apply filters & refinements, transform JSON) and on the specification of the executed operations (pre-process, postprocess) contained in the configuration document.

### 3.2. Configuration document

The configuration of the REST API is specified using a *hash-format* file (extension: `.hf`). The hash-format syntax, shown in Listing 1, is based on Markdown. A hash-format document includes several key-value pairs introduced by a hashtag, where the token flagged with the hashtag defines the name of a field and the following text is a Markdown content acting as a value associated with that field.

A RAMOSE configuration document includes two main sections, as shown in Listing 2. The first one contains general metadata and mandatory information about the REST API, and the other one includes a description of all the operations exposed by the REST

```
#<field_name_1> <field_value_1>
#<field_name_1> <field_value_2>
#<field_name_3> <field_value_3>
...
#<field_name_n> <field_value_n>
```

Listing 1. The hash-format syntax.

```
#url <api_base>
#type api
#title <api_title>
#description <api_description>
#version <version_number>
#endpoint <sparql_endpoint_url>
...

#url <operation_url_1>
#type operation
#sparql <sparql_query_1>
...

#url <operation_url_2>
#type operation
#sparql <sparql_query_2>
...
```

Listing 2. An excerpt of the structure of a RAMOSE configu-
ration document, organised in two sections: the one with in-
formation about the API (in *italic* in the listing), and the other
describing all the operations that the API exposes.

Table 1. The key-value pairs containing general information about
the API.

| #<field> <value> | Description |
|---|---|
| #url <api_base> | The base URL of the API (e.g. "/api/v1") |
| #type api | The section type – only "api" is allowed |
| #base <base_url> | The base URL of the webpage from which the API is available (e.g. "https://w3id.org/oc/index") |
| #method <get\|post> | The HTTP methods supported to send the request to the SPARQL endpoint, that can be "get", "post", or both |
| #title <api_title> | The title or name of the API |
| #description <api_description> | A textual description of the API |
| #version <version_number> | The textual string defining the version of the API |
| #license <license> | The textual string defining information about the licenses associated to the API, the data it returns, etc. |
| #contacts <contact_url> | The contact information for the API. |
| #endpoint <endpoint_url> | The SPARQL endpoint URL to query |
| #addon <addon_file_name> | The path of a Python file implementing functions that can be called in the preprocessing and postprocessing steps of each operation |

API. As examples, the RAMOSE configuration
documents we use in OpenCitations are available at
https://github.com/opencitations/api.

Table 1 lists all the fields used in the first section of
the configuration document to describe the REST API,
while Table 2 lists all the fields used to define all the
operations that may be included in the second section
of the configuration file. In both sections, `#url` must
be always the first field of each block.

### 3.3. Filters and refinements

RAMOSE implements optional filters and refine-
ment mechanisms on the results returned by the API.
These can be specified as HTTP parameters (i.e.
"?<param1>=<value1>&<param2>=<value2
>&...") in the API call URL.

These filters and refinement mechanisms work in-
dependently from the configuration file, the SPARQL
endpoint specified in it, and the scope of the RDF data
available. They provide common and advanced filter-
ing, sorting, and manipulative functionalities that can
be used with any result set returned by the API. The
operations that can be used are described as follows.

#### 3.3.1. Excluding rows with empty data

**Parameter: `require=<field>`.** All the rows that
have an empty value in the field `<field>` specified
as the value of the parameter are removed from the re-
sult set. E.g. `require=creation` removes all the
rows that do not have a value specified in the field
`creation`.

#### 3.3.2. Filtering rows

**Parameter: `filter=<field>:<opera-
tor><value>`.** Only the rows complying with the fil-
ter specified (i.e. `<field>:<opera-
tor><value>`) are considered in the result returned
by the API call. The term `<operator>` is not man-
datory.

If `<operator>` is not specified, `<value>` is in-
terpreted as a regular expression – e.g. `fil-
ter=creation:^20.+` returns the row in which
the value specified in the field creation starts with "20"
and it is followed by one or more characters. Other-
wise, if `<operator>` is specified, the value of
`<field>` of each row is compared with `<value>`
by means of the specified `<operator>`, that may as-
sume the following values: "=", "<", and ">". The
comparison will be done according to the type associ-
ated to the field under consideration, as specified in
`#field_type` (see Table 2). For instance, `fil-
ter=creation:>2016-05`, with the value type
specified as a datetime, returns all the rows that have
a date greater than 1 May 2016.

Table 2. The key-value pairs defining each operation of the API. All the fields accompanied with an "`[O]`" are optional in the configuration file.

| #`<field>` `<value>` | Description |
|---|---|
| #`url`<br>`<operation_url>` | The URL of the operation. It may contain zero or more parameters name between {…} (e.g. "`/cita-tions/{doi}`") |
| #`type operation` | The section type – only "operation" is allowed |
| #`<param>`<br>`<type>`(`<regex>`)<br>`[O]` | The shape (type and textual form) an input parameter of the operation must have (e.g. "`str(10\..+)`"). Possible types are strings ("`str`", which is the default value), integers ("`int`"), floating numbers ("`float`"), durations ("`duration`"), and date times ("`datetime`"). The regular expression is used to match the value of the parameter from the URL. |
| #`preprocess`<br>`<functions>` `[O]` | The Python functions used to pre-process the input parameters. One can specify one or more functions separated by "`-->`" which must take in input the name of one or more parameters (separated by a comma) between parenthesis, (e.g. "`lower(doi) --> encode(doi)`"). The output of a function is taken as input by the following functions. |
| #`postprocess`<br>`<functions>` `[O]` | The Python functions used to postprocess the results returned after the execution of the SPARQL query. One can specify one or more functions separated by "`-->`" which must take in input the name of zero or more variables (separated by a comma) returned by the SPARQL query between parenthesis (e.g. "`decode_doi(citing, cited)`") |
| #`method <get\|post>` | The HTTP method used to send the request to the SPARQL endpoint for this operation (that can be either "`get`" or "`post`") which must be compliant with those specified in the first section of the configuration file (as shown in Table 1) |
| #`description`<br>`<op_description>` | A textual description of the operation |
| #`field_type`<br>`<var_type_list>` | A list of types of the variables that will be returned by executing the operation, separated by space ("`<type1>(<var1>) <type2>(<var2>) …`") accompanied by their type – e.g. "`str(oci) datetime(creation) duration(timespan)`". Possible types are strings ("`str`", which is the default value), integers ("`int`"), floating numbers ("`float`"), durations ("`duration`"), and date times ("`datetime`") |
| #`call`<br>`<ex_request_call>` | The URL of an example of an API call (e.g. "`/citations/10.1108/jd-12-2013-0166`") |
| #`output_json`<br>`<ex_response>` | An example in JSON format of the results expected by the execution of the example call |
| #`sparql`<br>`<sparql_query>` | The SPARQL query to perform on the specified SPARQL endpoint. The query may include any input parameter of the operation between "`[[...]]`" (e.g. "`[[doi]]`") which is replaced with its current value before calling the SPARQL endpoint |

### 3.3.3. Sorting rows

**Parameter: `sort=<order>(<field>)`.** Sort in ascending (`<order>` set to "asc") or descending (`<order>` set to "desc") order the rows in the result set according to the values in `<field>`. For instance, `sort=asc(citing)` sorts all the rows according to the value specified in the field `citing` in ascending order.

### 3.3.4. Formatting results

**Parameter: `format=<type>`.** The final table is returned in the format specified in `<type>` that can be either "csv" (see Listing 3) or "json" (see Listing 4). For instance, `format=csv` returns the final table in CSV format. It is worth noting that this parameter takes priority over the format type specified in the "Accept" header of the HTTP request. Thus, if the header of a request to the API specifies `Accept: text/csv` and the URL of such request includes `format=json`, the final table is returned in JSON. This method allows developers (or users that may access data via browser) to override header requests.

```
citing,cited
10.3233/ds-190019,10.1108/jd-12-2013-0166
10.3233/sw-160224,10.1108/jd-12-2013-0166
…
```

Listing 3. A result set returned by RAMOSE in CSV format.

```
[
  {
    "citing":"10.3233/ds-190019",
    "cited":"10.1108/jd-12-2013-0166"
  },
  {
    "citing":"10.3233/sw-160224",
    "cited":"10.1108/jd-12-2013-0166"
  },
  …
]
```

Listing 4. The same result set returned by RAMOSE shown in Listing 3, but in JSON format.

```
[
  {
    "citing":"10.3233/ds-190019",
    "cited":["10.1108","jd-12-2013-0166"]
  },
  {
    "citing":"10.3233/sw-160224",
    "cited":["10.1108","jd-12-2013-0166"]
  },
  …
]
```

Listing 5. The same result in JSON shown in Listing 4, transformed according to the rule `array("/",cited)`, which splits the string value of the field `cited` according to the separator `/` and organises the resulting strings into a list.

### 3.3.5. Transforming JSON results

**Parameter:**
`json=<op>("<sep>",<field>,<new_field_1>,<new_field_2>,...).` When the JSON format is requested in the data return (see previous subsection), it is possible to transform each key-value pair of the final JSON according to the rule specified. Two possible operations `<op>` can be specified: "array" and "dict".

If `<op>` is set to "array", the string value associated with the key `<field>` is converted into an array by splitting the various textual parts at locations identified by means of the separator `<sep>`. For instance, considering the JSON shown in Listing 4, the execution of `array("/",cited)` returns the JSON shown in Listing 5.

Instead, if `<op>` is set to "dict", the value associated with the key `<field>` is converted into a JSON object by splitting the various textual parts using the separator `<sep>` and by associating each of these split strings to new fields specified by the keys `<new_field_1>`, `<new_field_2>`, etc. The number of newly specified keys corresponds to the number of splits to be applied to the value. For

```
[
  {
    "citing":{"prefix":"10.3233",
              "suffix":"ds-190019"},
    "cited":["10.1108","jd-12-2013-0166"]
  },
  {
    "citing":{"prefix":"10.3233",
              "suffix": "sw-160224"},
    "cited":["10.1108","jd-12-2013-0166"]
  },
  …
]
```

Listing 6. The same result in JSON shown in Listing 5, transformed according to the rule `dict("/",citing,prefix,suffix)`, which splits the string value of the field `citing` according to the separator `/` and organises the resulting strings into a JSON object with the new field labels `prefix` and `suffix`.

```
[
  {
    "citing":{"prefix":"10.3233",
              "suffix": "ds-190019"},
    "cited":[
      {"one":"1","two":".1108"},
      {"one":"jd-12-2","two": "13-0166"}
    ]
  },
  {
    "citing":{"prefix":"10.3233",
              "suffix": "sw-160224"},
    "cited":[
      {"one":"1","two":".1108"},
      {"one":"jd-12-2","two": "13-0166"}
    ],
  },
  …
]
```

Listing 7. The same result in JSON shown in Listing 6, transformed according to the rule `dict("0",cited,one,two)`, which splits each string value of the list in the field `cited` according to the separator `0` and organises the resulting strings into a JSON object according to the new fields `one` and `two`.

instance, considering the JSON shown in Listing 5, the execution of `dict("/",citing,prefix,suffix)` returns the JSON shown in Listing 6.

It is worth mentioning that, in case the value of the field has already been converted to a list of strings, the "dict" operation still works, and will be applied to all the strings contained in such a list. For instance, considering the JSON shown in Listing 6, the execution of `dict("0",cited,one,two)` returns the JSON shown in Listing 7.

### 3.3.6. Application of the filters and refinement mechanisms

In an API call, it is possible to specify one or more parameters of the same kind if you want to run the same filter and/or refinement multiple times. For instance, `require=citing&require=cited` excludes from the result all the rows that have unspecified the value of either the field `citing` or the field `cited`.

The order in which each parameter of the *same type* of a filter/refinement is run by RAMOSE depends on the order in which it is specified in the URL. However, the order of execution of the types of filter/refinement do not follow the actual order in the URL of the API call. Instead, RAMOSE first processes `require`, then `filter`, which is followed by `sort`. Then it applies `format` and, if the requested format is JSON, it finally executes `json`.

### 3.4. Run and deploy RAMOSE

There are three ways to run RAMOSE. First, one can use its command line interface (CLI) to execute. Second, it can be executed directly within a web server. Finally, it can be used directly within a Python code by using its main class, i.e. `APIManager`. These possibilities are described in the following subsections.

### 3.4.1. Command line interface (CLI)

RAMOSE can be run via CLI by specifying one or more configuration documents (parameter `-s`) and the operation to call (parameter `-c`), composed by concatenation of the API base URL with the operation URL, plus the wanted parameters for filtering and refining if needed. Also, it can take as input additional optional parameters (a) to specify the format of the output (parameter `-f`, JSON being the default), (b) to specify the name of the file in which to store the output (parameter `-o`, the output is printed in the shell output stream if a filename is not specified), and (c) to specify the method to use for the API request (parameter `-m`, GET being the default). The template of a CLI call of RAMOSE is shown as follows:

```
python ramose.py
    -s <conf_files>
    -c <api_base><operation_url>?<params>
    -f <csv|json>
    -o <output_name>
    -m <get|post>
```

RAMOSE can also create a HTML documentation of the API described in a configuration file. Specifically, the HTML documentation is requested by using the `-d` parameter, which can be stored in a file (parameter `-o`, as shown before) and, if needed, an additional CSS file can be specified to customise the layout of the document (parameter `-css`). The template of a CLI call of RAMOSE to generate the documentation is shown as follows:

```
python ramose.py
    -s <conf_file>
    -d
    -o <output_name>
    -css <css_file_path>
```

### 3.4.2. Web server

RAMOSE can also be used within a web server which is instantiated by using the parameter `-w` specifying the IP address of the host and the related port separated by ":" (e.g. `127.0.0.1:8080`). RAMOSE uses Flask to run the web server on the specified host machine. To deploy the REST API, one can use the following command:

```
python ramose.py
    -s <conf_files>
    -w <host:port>
    -css <css_file_path>
```

The Web API application can be accessed via a browser at the host and port specified (e.g. `http://127.0.0.1:8080`) and includes a basic dashboard for tracking API calls (available at `http://<host>:<port>`), and a documentation of the REST API (available at `http://<host>:<port>/<api_base>`).

### 3.4.3. Python classes

The Python class `APIManager` implements all the functionalities made available by RAMOSE. The signature of this class is as follows:

```
APIManager(conf_files)
```

The constructor of the class takes as input a list of API configuration files defined according to the hash format, and makes all the operations they define available for calling using the method `get_op` with the following signature:

```
get_op(op_complete_url)
```

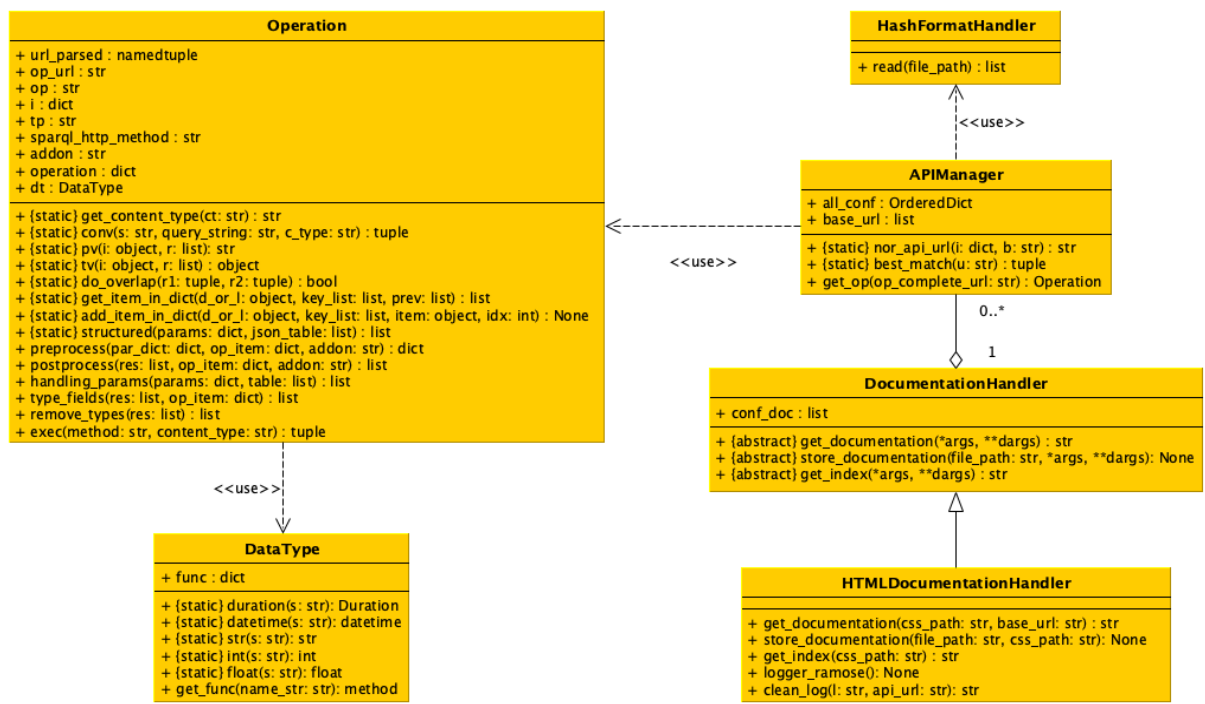

Figure 3. The UML class diagram of all the Python classes implementing RAMOSE.

This method takes as input a string containing the complete URL of the operation to execute, i.e. `<api_base>` plus `<operation_url>` such as `"/api/v1/citations/10.1108/JD-12-2013-0166"`, and returns an object of type `Operation` which implements the specific API call defined by the input URL. Thus, such an instance of the class `Operation` enables one to execute it by calling the method `exec` with the following signature:

```
exec(method, content_type)
```

This method takes as input the string describing the HTTP method to use to call the SPARQL endpoint (either `"get"` or `"post"`) and the content type (i.e. the format) of the result returned by the call (either `"csv"` or `"json"`). The method returns a tuple of two items. The first item contains the status code of the HTTP response, while the second item contains the string of the results in the requested format. The UML class diagram of all the classes implementing RAMOSE is shown in Figure 3.

## 4. Use of REST APIs in OpenCitations

To date, RAMOSE has been principally adopted by the OpenCitations organisation to serve its datasets to a variety of stakeholders, including both Semantic Web practitioners and scholars in research fields that are not familiar with Semantic Web technologies, e.g. Scientometrics. This case study gives us the opportunity to sketch a preliminary evaluation of the software in order to propose a mature, flexible solution to the community and to foster its reuse as based on the evidence of actual benefits for RDF data providers.

Specifically, we aim to understand the potential benefit that the introduction of REST APIs - created by means of RAMOSE - can bring for access to and reuse of data stored in RDF. We analysed the logs of the requests to OpenCitations services between January 2018 and March 2020 to identify trends in data access strategies. This period was particularly meaningful, since the first REST API made available by OpenCitations was released in June 2018, before

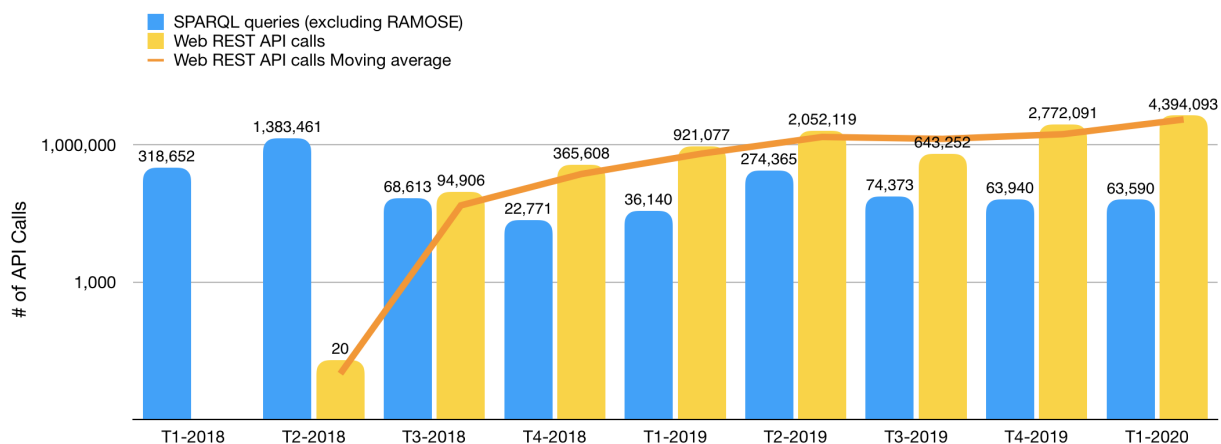

Figure 4. The number of requests received by the OpenCitations SPARQL endpoints vs. the calls to the OpenCitations REST APIs between January 2018 and March 2020 – listed by trimester. The orange line represents the moving average of the number of Web REST API calls. Note that the vertical axis has a logarithmic scale.

which our data was available only through SPARQL endpoints.

While this paper represents the first proposal of RA-MOSE to the community for direct use in accessing other RDF triplestores, several stakeholders have requested systematic access to the OpenCitations REST APIs. We give an account of applications and services that systematically rely on OpenCitations REST APIs created via RAMOSE.

### 4.1. Users' queries analysis

As mentioned above, the logs allow us to understand to what extent the introduction of the REST API has changed the way users interact with OpenCitations data. However, although OpenCitations implements an open REST API system, it does not track users (e.g. by means of API keys) and users' IP addresses are currently masked by proxies, hence the impact of the introduction of RAMOSE REST APIs cannot be directly measured in terms of unique users. Being aware of a potential bias in results, we consider the number of queries (either directly against the SPARQL endpoint or to a Web REST API) as the unit of measure for showing trends in the two strategies for data access.

Specifically, we first analysed whether Web REST API operations implemented by OpenCitations are designed in a way that satisfies a representative number of users' queries, to assume that the types of queries performed directly against the SPARQL endpoint can be reasonably compared (one-to-one) with REST API calls.

We collected all the SPARQL queries performed by users against the SPARQL endpoints of the OpenCitations Corpus and of the OpenCitations Indexes over a 3-year time span (from January 2018 to March 2020), removed common variable terms (e.g. prefixes, limit, offset) and normalised query variables to placeholder strings (e.g. specific DOI requested in a query is substituted with the generic term `"string"`). We pruned all the queries that had less than four query variables and those that had been performed less than ten times, to remove the long tail of queries that may have been the results of failed attempts, mistakes and so on. We then aggregated queries by using the affinity propagation method (i.e. a method that does not require one a priori to specify the number of final clusters), obtaining ten clusters of queries, for each of which an exemplar query is provided. We finally matched exemplar queries with the five queries underlying OpenCitations API calls by measuring their string similarity. We selected a similarity score (see *token set* ratio, https://github.com/seatgeek/fuzzywuzzy) that allows us to measure whether the same triple patterns appear in the matching query regardless their order.

In Table 3 we illustrate the results of the comparison, including for each cluster the name of the API call having the best match with respect to the exemplar query, and the similarity score expressed in percentage.

The results show that, with an average confidence of ∼80%, users' SPARQL queries are covered by two of the main API calls designed by OpenCitations, namely: *metadata*, a method to retrieve bibliographic metadata of resources and their references, and *coauthorship*, a method to retrieve the network of authors given a set of documents (see https://opencitations.net/api/v1 and https://opencitations.net/index/api/v1 for a complete description of the API calls).

Table 3. Similarity between clusters of users' queries and API calls.

| Cluster | Matched API call name | Similarity (%) |
|---------|----------------------|----------------|
| 1 | metadata | 71 |
| 2 | metadata | 88 |
| 3 | metadata | 79 |
| 4 | coauthorship | 79 |
| 5 | coauthorship | 93 |
| 6 | metadata | 90 |
| 7 | metadata | 99 |
| 8 | metadata | 94 |
| 9 | coauthorship | 86 |
| 10 | metadata | 65 |

Secondly, we compared the number of total SPARQL queries made against the OpenCitations SPARQL endpoints (excluding those coming from RAMOSE) with the number of all the REST API calls received in the same period. The results, split by trimester for the sake of readability, are shown in Figure 4.

In Figure 4, the blue bars show usage employing the OpenCitations SPARQL endpoints directly or using the other available non-API services, while the yellow bars show access using the APIs created with RAMOSE. The increase in overall usage of OpenCitations datasets following the introduction of the APIs may be attributed both to the increased ease of access to OpenCitations data that these APIs make possible, thus attracting use by people unfamiliar with SPARQL, and to the launch in June 2018 of COCI [11], which for the first time made available through OpenCitations the hundreds of millions of citations derived from open references at Crossref.

While there is some fluctuation in the quarterly figures, there has been a significant increase in the average number of API calls since T3-2018, and a significant declining trend in the interaction with the other SPARQL services. That trimester could be considered the turning point, since at that time several developers with no or limited expertise with Semantic Web technologies started to build prototype applications using the OpenCitations data newly available via the REST API. In that trimester, the total number of accesses to the REST API was 138% of the number of accesses to the other SPARQL services. In subsequent trimesters, the use of the original SPARQL services decreased substantially to become stable at about 22,000 requests per month, while the number of REST API calls increased dramatically, reaching a total number of 4,394,093 calls in T1-2020, more than 65 times the number of SPARQL requests.

These figures point to substantial benefits from the adoption of RAMOSE with regards to increasing user interaction with OpenCitations data. Such results are corroborated by results of the preliminary comparison, which allow us to validate the conclusion that a decrease of the number of SPARQL queries in favour of a higher number of REST API calls can be associated with an actual increase of data engagement – rather than being due to bad API design choices and a consequent inefficient increase of the number of API calls.

The source code for reproducing the preliminary analysis is available at http://github.com/opencitations/ramose/tree/master/eval. All the data used in this evaluation are available on Zenodo [13].

### 4.2. Current uptake of RAMOSE

The flexibility of RAMOSE enabled the simple creation of additional REST APIs for each of the new datasets released by OpenCitations. From the first REST API released in June 2018, three other REST APIs have been released, as described at http://opencitations.net/querying, with all the configuration documents being available at https://github.com/opencitations/api.

Other REST APIs based on RAMOSE, for services external to OpenCitations, have been developed to address specific tasks. For instance, during the Hack Day of the 2018 Workshop on Open Citations (https://workshop-oc.github.io/2018/), we developed an exemplar REST API service (still available at http://opencitations.net/wikidata/api/v1) to return scholarly metadata from the Wikidata SPARQL endpoint (https://query.wikidata.org). This REST API has been used by the citation network visualisation tool VOSviewer [14] (https://www.vosviewer.com) to display the citation network within more than 5,000 papers in the Wikidata Zika Corpus (https://twitter.com/ReaderMeter/status/1037349669335126016).

In addition to VOSviewer, the REST APIs developed by OpenCitations using RAMOSE have been extensively used in several other software and data services. Those of which we are aware are Citation Gecko (https://citationgecko.com), OpenAccess Helper (https://www.oahelper.org), DBLP (https://dblp.uni-trier.de), CiteCorp (https://github.com/ropensci-labs/citecorp), and Zotero (https://github.com/zuphilip/zotero-open-citations). Our interactions with the developers of those services have been instrumental in guiding the development of the facilities that RAMOSE makes available, including the filters and refinement mechanisms that have demonstrated their usefulness is several scenarios, and have led, for example, to the adoption of JSON as the default data format returned by RAMOSE so as to meet to the input requirements of VOSviewer.

## 5. Related works

In the past, several tools, in particular REST APIs on top of SPARQL endpoints, have been developed to leverage RDF data served through SPARQL query interfaces, often employing bespoke solutions tailored to their data, such as the DBpedia REST APIs (https://wiki.dbpedia.org/rest-api) and the Europeana Search API (https://pro.europeana.eu/page/search).

Among works that are closer to what RAMOSE provides, the following deserve specific mention.

BASIL [15] is a cloud platform that supports sharing and reusing of SPARQL queries, and automatically generates Web APIs from those, which can be easily embedded into users' applications. Moreover, it allows one to reuse results as HTML snippets, called "views". While pre-processing operations are possible, the only way to undertake post-processing operations is separately to implement ad-hoc procedures on the returned results. BASIL runs using Java and requires the installation and configuration of a MySQL server on the running machine.

Another important tool in this category is grlc (http://grlc.io/) [16], a lightweight server that translates on the fly to Linked Data API calls SPARQL queries stored in a GitHub repository, in a local filesystem, or listed at a URL. The idea behind grlc is to implement an API mapping along with the use of SPARQL decorators which extends the original queries with other generated metadata which add extra functionalities to the APIs. In order to make this happen, the specified archive must contain a collection of SPARQL queries as `.rq` files and include the decorators as comments inside each `.rq` file. With grlc the pre-/post-processing operations are defined as decorators, and each API call can point to a different SPARQL endpoint by specifying the decorator "endpoint" before the SPARQL query.

A useful add-on to integrate with the grlc, suggested by its authors, is SPARQL Transformer [17]. This tool allows one to simplify the JSON outputs of a SPARQL query by re-shaping and simplifying the final JSON schema. SPARQL Transformer relies on a single JSON object to define which data should be extracted from the endpoint and what shape should they assume. Although this approach refines the final output, using it alone does not allow one to perform custom operations on the returned results (e.g. data normalisation or cleansing), which are delegated to separated post-processing operations, e.g. using the grlc features, or ad-hoc functions. SPARQL Transformer is written in JavaScript and can be imported and integrated in an HTML module.

Another recently proposed framework is the Ontology-Based API (OBA) [24]. In contrast with the other tools discussed above, OBA creates a REST API service starting from an ontology (specified in OWL), and it performs queries only on one SPARQL endpoint, which must expose RDF data modelled according to the ontology used to create the API. OBA automatically generates a series of SPARQL queries templates to execute CRUD (create, retrieve, update, delete) operations through the REST API generated, and gives the possibility to define other API operations based on custom SPARQL queries. This customisation is provided using the same strategy adopted in grlc.

In Table 4 we mention the tools discussed above except the last one and highlight some of their meaningful features. We have also included RAMOSE in the table, to enable a comparison with other tools. The table header contains the tools analysed. The first column lists the main features retrieved by analysing the tools and organises them in three macro groups:

1.  Tool characteristics: the technical characteristics of the tool including the programming language used for its creation, its license, its requirements, and how to run it.
2.  Building the REST API service: what modules and configuration files should be defined to create the REST API service by using the tool, and what inputs/data are required.
3.  REST API service: the characteristics of the REST API service produced using the tool, introducing the main aspects of its API operations, API parameters, documentation, and other features worth of mentioning.

Table 4. A comparison between RAMOSE and other tools used to create RESTful APIs over RDF triplestores. The first two columns define the features considered in the analysis. The features are classified under three macro groups: "Tool characteristics", which describe the technical characteristics of the tool, "Building the REST API service using the tool", i.e. the modules/configuration files and the values needed for the creation of the REST API service, and "REST API service", which illustrates the features of the created REST API service.

| | | RAMOSE | grlc | Basil | OBA |
|---|---|---|---|---|---|
| **Tool characteristics** | *Language* | Python | Python | Java | Java |
| | *License* | ISC License | MIT License | N/A | Apache License |
| | *Running requirements* | + Python should be installed on the running machine | + Java should be installed on the running machine<br>+ Have a MySQL server.<br>+ A database on MySQL running a specific list of queries<br>+ A configuration file in .ini format, which defines the connection parameters.<br>+ A log4j2 configuration file for logging permissions. **(optional)** | + Java should be installed on the running machine<br>+ Docker should be installed on the running machine |
| | *Running interface* | + CLI<br>+ Web server<br>+ Python code | + Its web service (grlc.io service) | + CLI | + Python<br>+ Javascript client |
| **Building the REST API service using the tool** | *Values (required or optional)* | + A URL to the SPARQL endpoint<br>+ Custom SPARQL queries<br>+ General metadata about the REST API service to create | + A URL to the SPARQL endpoint<br>+ Custom SPARQL queries<br>+ General metadata about REST API service to create | + A URL to the SPARQL endpoint<br>+ Custom SPARQL queries<br>+ Views: an alternative presentation of the API results based on a template or script, e.g. HTML representation. **(optional)** | + A URL to the SPARQL endpoint<br>+ Ontology network (specified in OWL)<br>+ Custom SPARQL queries **(optional)** |
| | *Module/s (required or optional)* | + One configuration file in *hash-format* | + A list of files in *rq* format (each file specifies a different API operation).<br>+ Configuration files could be loaded from a GitHub repository, a local storage, or a spec file accessible on the web | | + A configuration file in *.yaml* format.<br>+ A list of files in *rq* format (each file specifies a different API operation) **(optional)** |
| **REST API service** | *Parameters* | + Filters<br>+ Ordering<br>+ Result formats (CSV or JSON)<br>+ JSON transformation | + Specify a page (in case a maximum number of results are specified)<br>+ Parameters used as variables inside the SPARQL queries definition (following a particular convention)<br>+ Parameters values could be restricted to a set of permitted options | + All the available parameters are used as variables inside the SPARQL queries definition (following a particular convention) | + All the available parameters are used as variables inside the SPARQL queries definition (following a particular convention) |
| | *Operations* | + Via GET or POST requests<br>+ Can perform a list of pre-/post-processing methods to apply on respectively the input of the API operation and output of the SPARQL query executed by that operation | + Via GET or POST requests<br>+ Transformation/re-engineering in JSON of the output returned by the SPARQL query executed by that operation | | + Via GET, POST, PUT or DELETE requests<br>+ Supports authorization when using POST, PUT and DELETE methods |
| | *Documentation* | + Own Web-based documentation<br>+ Automatically built according to the values specified in the configuration file | + Swagger-based documentation<br>+ Automatically built according to the values specified in the configuration file | + Swagger-based documentation | + Swagger-based documentation<br>+ Automatically built following the annotations in the input ontology |
| | *Others* | + Perform queries on different SPARQL endpoints only through SPARQL federation<br>+ Import Python files to expose functions for pre- and post-processing methods | + Perform queries on different SPARQL endpoints<br>+ Generation of provenance in PROV of both the Git repository history and grlc's activity additions | + Perform queries on different SPARQL endpoints<br>+ Creates new API operations using HTTP PUT requests | + Performs queries only on one SPARQL endpoint (since it is assumed to be modelled according to the ontology used to create the API)<br>+ It generates automatic SPARQL queries templates to handle CRUD-based APIs |

Among the several differences between RAMOSE and the other tools, we want to highlight the following two:

- all the tools analysed generate a web-based front-end documentation based on Swagger (https://swagger.io/). In RAMOSE, we propose a different approach with an alternative representation, the main purpose of which is to be as far as possible clear to those who are not software programmers;
- the parameters of RAMOSE are based on pre-defined options/functions, and it is possible to customize pre-processing and post-processing operations. The other tools define the REST API parameters based on variables that form part of the SPARQL queries (which follow a specific convention).

In addition to the tools mentioned above, it is worth mentioning also the approach proposed in [18]. In this work Schröder *et al.* present a generic approach to convert any given SPARQL endpoint into a path-based JSON REST API. This work focused mostly on simple CRUD-based workflows. The idea behind this approach is to build API paths that follow RDF triple patterns, e.g. the call `/class/dbo:Coun-try/dbr:Germany` returns a JSON object for the specified entity (i.e. the DBpedia resource representing Germany). Despite being very intuitive for Semantic Web practitioners, it does not help adopters who are not acquainted both with RDF knowledge organisation and the scope of the dataset at hand.

Other SPARQL editor interfaces have been published in the past, with the aim of assisting users in querying against SPARQL endpoints by means of a user-friendly GUI, e.g. YASGUI [19]. Such tools are meant to allow users to perform exploratory queries, but do not offer means to programmatically access data. Moreover, these are usually hard to use by users with no knowledge of SPARQL.

Another class of tools include WYSIWYG Web applications for searching and browsing RDF data by hiding the complexity of SPARQL. Such tools include general-purpose RDF search engines and GUI interfaces, such as Pubby (http://wifo5-03.informatik.uni-mannheim.de/pubby/), LodView [20], our own search interface OSCAR [21], Scholia [22], Elda (http://www.epimorphics.com/web/tools/elda.html), and BioCarian [23].

Finally, some frameworks have been developed in the past in order to provide high-level interfaces to interact with RDF data through a web server. Among these, it is worth mentioning the Python Linked Data

API (pyLDAPI, https://github.com/RDFLib/pyLDAPI), which is Python module that can be added into a Python Flask installation to handle requests and return responses in a manner consistent with Linked Data principles of operation.

## 6. Conclusions

In this article, we have introduced RAMOSE, the RESTful API Manager Over SPARQL Endpoints. RAMOSE is an open-source Python software development that allows one to create Web REST API interfaces to one or more SPARQL endpoints by editing a configuration file in Markdown-like syntax, automatically generating documentation and a web server for testing and monitoring purposes. This generic software, which is freely available on GitHub and citable via Zenodo [25], can be used over any SPARQL endpoint by creating a configuration text file. We have illustrated all the features that RAMOSE implements and we have presented the analysis of our motivating scenario, namely the dramatic increase in usage of OpenCitations data demonstrated by our access logs, to demonstrate the benefit that such a tool has brought to OpenCitations in terms of user interaction with its data. We commend the use of RAMOSE to others wishing to expose their own RDF data via a REST API.

In the future, we aim at extending RAMOSE with missing CRUD (Create, Read, Update, Delete) operations to fully support Semantic Web developers when interacting with the triplestores they own. Secondly, we want to enhance RAMOSE capabilities and support Web developers in interacting with other types of data sources, such as JSON, XML, CSV data dumps and relational databases, so that it will be possible to leverage the same software solution over different data sources. Lastly, alongside the method for importing CSS templates, we will provide methods to import custom HTML templates.

## References

[1] R. T. Fielding, 'REST APIs must be hypertext-driven', *Untangled musings of Roy T. Fielding*, Oct. 20, 2008. Accessed: Nov. 12, 2020. Available: https://roy.gbiv.com/untangled/2008/rest-apis-must-be-hypertext-driven.

[2] C. Ogbuji, 'SPARQL 1.1 Graph Store HTTP Protocol', World Wide Web Consortium, W3C Recommendation, Mar. 2013. Accessed: Nov. 12, 2020. Available: https://www.w3.org/TR/sparql11-http-rdf-update/.

[3] R. Cyganiak, D. Wood, and M. Krötzsch, 'RDF 1.1 Concepts and Abstract Syntax', World Wide Web Consortium, W3C Recommendation, Feb. 2014. Accessed: Nov. 12, 2020. Available: https://www.w3.org/TR/rdf11-concepts/.

[4] S. Harris and A. Seaborne, 'SPARQL 1.1 Query Language', World Wide Web Consortium, W3C Recommendation, Mar. 2013. Accessed: Nov. 12, 2020. Accessed: Nov. 12, 2020. Available: https://www.w3.org/TR/sparql11-query/.

[5] S. Peroni and D. Shotton, 'OpenCitations, an infrastructure organization for open scholarship', *Quantitative Science Studies*, vol. 1, no. 1, pp. 428–444, 2020, doi: 10.1162/qss_a_00023.

[6] S. Peroni, D. Shotton, and F. Vitali, 'One Year of the OpenCitations Corpus', in *The Semantic Web – ISWC 2017*, Cham, Switzerland, 2017, vol. 10588, pp. 184–192, doi: 10.1007/978-3-319-68204-4_19.

[7] S. Peroni and D. Shotton, 'The SPAR Ontologies', in *The Semantic Web – ISWC 2018*, Cham, Switzerland, 2018, Lecture Notes in Computer Science vol. 10842, pp. 119–136, doi: 10.1007/978-3-030-00668-6_8.

[8] G. Hendricks, D. Tkaczyk, J. Lin, and P. Feeney, 'Crossref: The sustainable source of community-owned scholarly metadata', *Quantitative Science Studies*, vol. 1, no. 1, pp. 414–427, 2020, doi: 10.1162/qss_a_00022.

[9] L. L. Haak, M. Fenner, L. Paglione, E. Pentz, and H. Ratner, 'ORCID: a system to uniquely identify researchers', *Learned Publishing*, vol. 25, no. 4, pp. 259–264, 2012, doi: 10.1087/20120404.

[10] S. Peroni and D. Shotton, 'Open Citation Identifier: Definition'. Figshare, 2019, doi: 10.6084/m9.figshare.7127816.

[11] I. Heibi, S. Peroni, e D. Shotton, 'The OpenCitations Index of Crossref open DOI-to-DOI citations', *Scientometrics*, vol. 121, n. 2, pp. 1213–1228, 2019, doi: 10.1007/s11192-019-03217-6.

[12] M. Saleem, S. Gábor, F. Conrads, S. A. C. Bukhari, Q. Mehmood, and A. Ngonga Ngomo. 'How representative is a sparql benchmark? an analysis of rdf triplestore benchmarks', in *WWW '19: The World Wide Web Conference*, pp. 1623-1633. 2019.

[13] M. Daquino, I. Heibi, S. Peroni, and D. Shotton, 'OpenCitations 2018-2020 requests: SPARQL endpoints vs REST APIs v2'. Zenodo, 2020, doi: 10.5281/ZENODO.3953068.

[14] N. J. van Eck and L. Waltman, 'Software survey: VOSviewer, a computer program for bibliometric mapping', *Scientometrics*, vol. 84, no. 2, pp. 523–538, 2010, doi: 10.1007/s11192-009-0146-3.

[15] E. Daga, L. Panziera, and C. Pedrinaci, 'BASIL: A Cloud Platform for Sharing and Reusing SPARQL Queries as Web APIs', in *ISWC-P&D 2015 - ISWC 2015 Posters & Demonstrations Track*, Aachen, Germany, 2015, CEUR Workshop Proceedings vol. 1486. Accessed: Nov. 12, 2020. Available: http://ceur-ws.org/Vol-1486/paper_41.pdf.

[16] A. Meroño-Peñuela and R. Hoekstra, 'grlc Makes GitHub Taste Like Linked Data APIs', in *The Semantic Web*, Cham, Switzerland: Springer, 2016, Lecture Notes in Computer Science vol. 9989, pp. 342–353, doi: 10.1007/978-3-319-47602-5_48

[17] P. Lisena, A. Meroño-Peñuela, T. Kuhn, and R. Troncy, 'Easy Web API Development with SPARQL Transformer', in *The Semantic Web – ISWC 2019*, Cham, Switzerland: Springer, 2019, Lecture Notes in Computer Science vol. 11779, pp. 454–470.

[18] M. Schröder, J. Hees, A. Bernardi, D. Ewert, P. Klotz, and S. Stadtmüller, 'Simplified SPARQL REST API: CRUD on JSON Object Graphs via URI Paths', in *The Semantic Web: ESWC 2018 Satellite Events*, Cham, Switzerland: Springer, 2018, Lecture Notes in Computer Science vol. 11155, pp. 40–45.

[19] L. Rietveld and R. Hoekstra, 'The YASGUI family of SPARQL clients', *Semantic Web*, vol. 8, no. 3, pp. 373–383, 2016, doi: 10.3233/SW-150197.

[20] D. V. Camarda, S. Mazzini, and A. Antonuccio, 'LodLive, exploring the web of data', in *Proceedings of the 8th International Conference on Semantic Systems - I-SEMANTICS '12*, New York, NY, USA: ACM, 2012, pp. 197–200, doi: 10.1145/2362499.2362532.

[21] I. Heibi, S. Peroni, and D. Shotton, 'Enabling text search on SPARQL endpoints through OSCAR', *Data Science*, vol. 2, no. 1–2, pp. 205–227, 2019, doi: 10.3233/DS-190016.

[22] F. Å. Nielsen, D. Mietchen, and E. L. Willighagen, 'Scholia, Scientometrics and Wikidata', in *The Semantic Web: ESWC 2017 Satellite Events*, Cham, Switzerland: Springer, 2017, Lecture Notes in Computer Science vol. 10577, pp. 237–259, doi: 10.1007/978-3-319-70407-4_36.

[23] N. Zaki and C. Tennakoon, 'BioCarian: search engine for exploratory searches in heterogeneous biological databases', *BMC Bioinformatics*, vol. 18, no. 1, Dec. 2017, doi: 10.1186/s12859-017-1840-4.

[24] D. Garijo and M. Osorio, 'OBA: An Ontology-Based Framework for Creating REST APIs for Knowledge Graphs', in *The Semantic Web – ISWC 2020*, Cham, Switzerland: Springer, 2020, Lecture Notes in Computer Science vol. 12507, pp. 48–64, doi: 10.1007/978-3-030-62466-8_4.

[25] S. Peroni and M. Daquino, 'opencitations/ramose: Release 1.0.2', Version 1.0.2 [software]. Zenodo. doi: 10.5281/zenodo.4585536